\documentclass[12pt,preprint,twocolumns]{emulateapj}
\usepackage{latexsym}
\usepackage{url}
\usepackage{color,ulem}
\usepackage[hypertex]{hyperref}
\usepackage{amssymb}
\usepackage{amsmath}
\usepackage{epsfig}
\usepackage{appendix}
\usepackage{wrapfig}
\singlespace
\def\HI{\hbox{\rm H~$\scriptstyle\rm I$}}

\def\HII{\hbox{\rm H~$\scriptstyle\rm II$}}
\def\nHI{{\rm HI}}
\def\nH{{\rm H}}
\def\nHII{{\rm HII}}
\def\nHe{{\rm He}}

\def\nHeII{{\rm HeII}}
\def\nHeIII{{\rm HeIII}}

\def\HeI{\hbox{He~$\scriptstyle\rm I$}}
\def\HeII{\hbox{He~$\scriptstyle\rm II$}}
\def\HeIII{\hbox{He~$\scriptstyle\rm III$}}

\def\HeIII{\hbox{He~$\scriptstyle\rm III$}}

\def\kmsmpc{\,{\rm km\,s^{-1}\,Mpc^{-1}}}

\def\uvunits{\,{\rm erg\,cm^{-2}\,s^{-1}\,Hz^{-1}\,sr^{-1}}}

\def\lumdens{\,{\rm erg\,s^{-1}\,Mpc^{-3}\,Hz^{-1}}}

\def\Lya{Ly$\alpha$}
\def\Lyb{Ly$\beta$}

\def\etal{{et al.\ }}

\def\spose#1{\hbox to 0pt{#1\hss}}
\def\lta{\mathrel{\spose{\lower 3pt\hbox{$\mathchar"218$}}
     \raise 2.0pt\hbox{$\mathchar"13C$}}}
\def\gta{\mathrel{\spose{\lower 3pt\hbox{$\mathchar"218$}}
     \raise 2.0pt\hbox{$\mathchar"13E$}}}
\def\ni{\noindent}

\lefthead{Madau and Haardt}
\righthead{Reionization by AGNs}
\submitted{}


\begin{document}

\title{Cosmic Reionization after Planck: Could Quasars Do It All?}

\author{Piero Madau\altaffilmark{1} and Francesco Haardt\altaffilmark{2,3}}
\affil{$^1\,$Department of Astronomy \& Astrophysics, University of California, 1156 High Street, Santa Cruz, CA 95064, USA\\
$^2\,$Dipartimento di Scienza e Alta Tecnologia, Universit\`a dell'Insubria, via Valleggio 11, 22100 Como, Italy\\
$^3\,$ INFN, Sezione Milano/Bicocca, P.za della Scienza 3, 20126 Milano, Italy}

\begin{abstract}
We assess a model of late cosmic reionization in which the ionizing background radiation arises entirely from high redshift quasars and other active 
galactic nuclei (AGNs). The low optical depth to Thomson scattering reported by the Planck Collaboration pushes the redshift of instantaneous reionization 
down to $z=8.8^{+1.7}_{-1.4}$ and greatly reduces the need for significant Lyman-continuum emission at very early times. 
We show that, if recent claims of a numerous population of faint AGNs at $z=4-6$ are upheld, and the 
high inferred AGN comoving emissivity at these epochs persists to higher, $z\gta 10$, redshifts, then active galaxies may drive the 
reionization of hydrogen and 
helium with little contribution from normal star-forming galaxies. We discuss an AGN-dominated scenario that satisfies a number of observational
constraints: the \HI\ photoionization rate is relatively flat over the range $2<z<5$, hydrogen gets fully reionized by $z\simeq 5.7$, 
and the integrated Thomson scattering optical depth is $\tau\simeq 0.056$, in agreement with measurements based on the \Lya\ opacity 
of the intergalactic medium (IGM) and cosmic microwave background (CMB) polarization. It is a prediction of the model that helium gets doubly 
reionized before redshift 4, the heat input from helium reionization dominates the thermal balance of the IGM after hydrogen reionization, 
and $z>5$ AGNs provide a significant fraction of the unresolved X-ray background at 2 keV. Singly- and doubly-ionized helium contribute about 13\% to 
$\tau$, and the \HeIII\ volume fraction is already 50\% when hydrogen becomes fully reionized. 
\end{abstract}


\section{Introduction}

The reionization of the all-pervading IGM marks a turning point in the history of structure formation in the universe. 
The details of this process reflect the nature of the first astrophysical sources of radiation and heating as well as the early thermodynamics of 
cosmic baryons, and continue to be the subject of considerable observational and theoretical efforts \citep[for a recent review, see][]{Robertson10}. 
Studies of resonant absorption in the spectra of distant quasars show that, while inhomogeneous hydrogen reionization may still be ongoing at
$z\sim 6$, it has fully completed by redshift 5 \citep[e.g.,][]{Becker15,McGreer15,Fan06}. It is generally agreed that the IGM is kept ionized 
by the integrated UV emission from active nuclei and star-forming galaxies, but there is still no consensus on the relative contributions of these 
sources as a function of epoch. The high ionization threshold (4 ryd) and small photoionization cross section of \HeII, the rapid recombination rate of 
\HeIII, and the fact that most hot stars lack 4 ryd emission, all delay the double ionization of helium. This is expected to be completed by 
hard UV-emitting quasars and other AGNs around the apparent peak of their activity at $z\approx 2.5$ \citep[e.g.,][]{McQuinn09,Haardt12}, 
later than the reionization of \HI\ and \HeI.  It is the traditional view that, at $z>3$, as the declining population of optically bright 
quasars makes an increasingly small contribution to the 1 ryd radiation background \citep[e.g.][]{Shapiro87}, massive stars in early galactic 
halos take over and provide the additional ionizing flux needed \citep[e.g.,][]{Madau99,Meiksin05,FG09}. 

While plausible, this ``two-components" picture for cosmic reionization is not necessarily correct, and warrants 
further investigation for validation. In particular, star-forming galaxies at $z>5$ can keep the Universe substantially ionized only if one 
extrapolates the steep UV luminosity function well below the observed limits and assumes a globally-averaged absolute escape fraction of Lyman-continuum (LyC)
radiation into the early IGM, $\bar f_{\rm esc}$, that exceeds 20\% \citep[e.g.,][]{Bouwens12,Finkelstein12,Haardt12}. Such 
leakage values are higher than typically
inferred from observations of luminous galaxies at $z\sim 3-4$ once contamination by foreground low-redshift interlopers 
is accounted for \citep{Vanzella12}. Despite significant efforts and the examination of hundreds of galaxies, there exists only a handful of 
robust detections as of today \citep[e.g.,][]{Siana15,Mostardi15}.

The Planck Collaboration (2015) has reported a new, smaller value, $\tau=0.066\pm 0.016$ of the integrated reionization optical depth 
from low-multipoles polarization, lensing, and high-multipole temperature CMB data. This corresponds to a sudden reionization event 
at $z=8.8^{+1.7}_{-1.4}$ and reduces the need for a large LyC background at very early times. Together with the recent claim by 
\citet{Giallongo15} (see also \citealt{Glikman11}) 
of a significant population of faint AGNs at $4<z<6.5$, these facts have prompted us to reassess a scenario in which quasars and active galaxies may 
actually dominate the cosmic reionization process at all epochs, with normal star-forming galaxies making only a negligible contribution due 
to their small leakages. We explore this intriguing possibility below, assuming a $(\Omega_M,\Omega_\Lambda,\Omega_b)=(0.3,0.7,0.045)$
flat cosmology throughout with $H_0=70\,\kmsmpc$.

\section{QSO Comoving Emissivity}

\begin{figure}[thb]
\centering
\includegraphics*[width=0.49\textwidth]{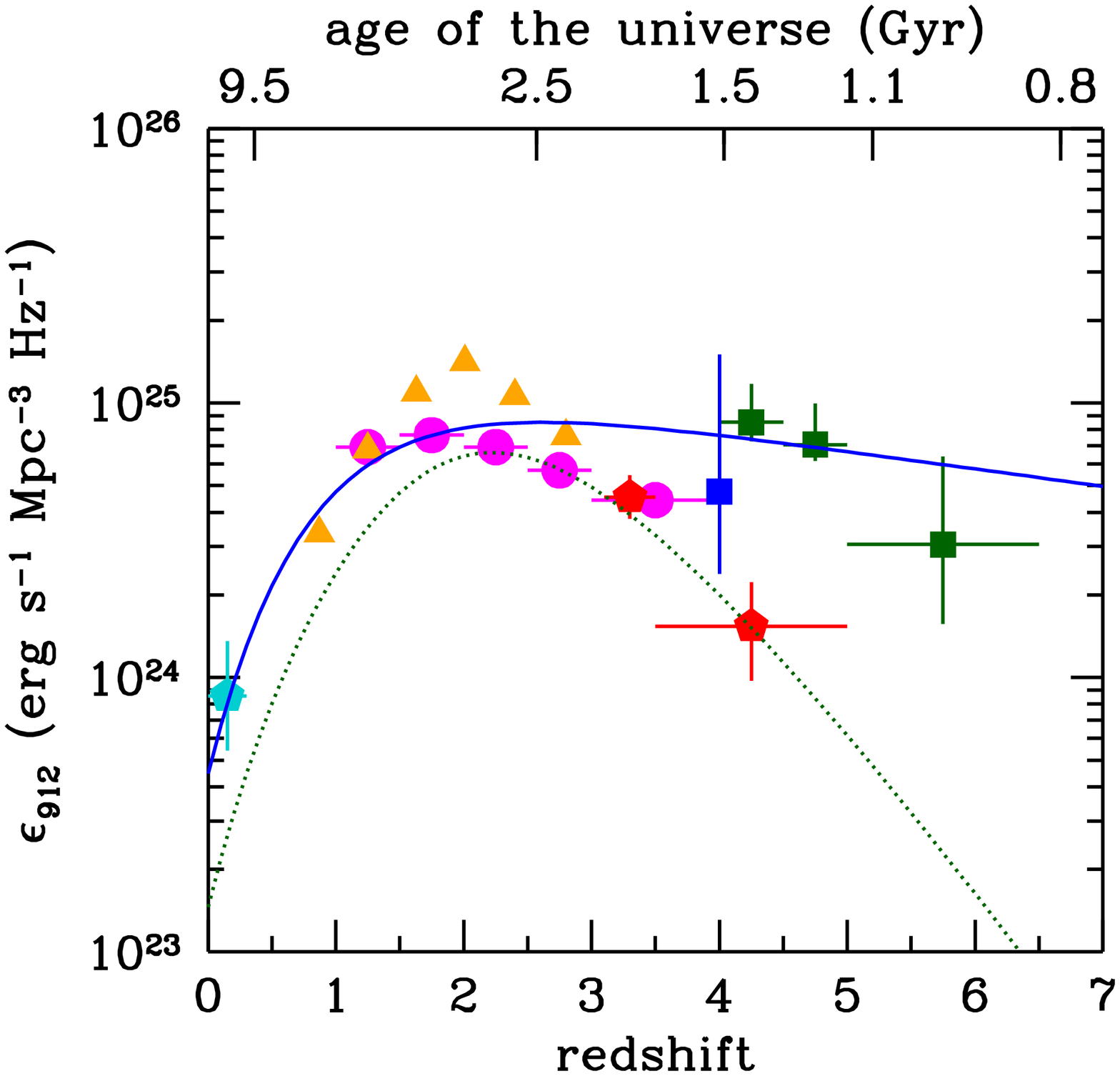}
\caption{\footnotesize The AGN comoving ionizing emissivity inferred from \citet{Schulze09} (cyan pentagon), 
\citet{Palanque-Delabrouille13} (orange triangles), \citet{Bongiorno07} (magenta circles), \citet{Masters12} (red pentagons),
\citet{Glikman11} (blue square), and \citet{Giallongo15} (green squares). The solid curve shows the functional form given in Equation
(\ref{eqemiss}). The LyC AGN emissivity of \citet{Hopkins07} is shown for comparison (dotted line).
See text for details.  
}
\label{fig1}
\vspace{+0.cm}
\end{figure}

Figure \ref{fig1} shows the inferrred quasar/AGN comoving emissivity at 1 ryd as a function of redshift. Our modeling is based on a 
limited number of contemporary, optically-selected AGN samples (see also \citealt{Khaire15} for a similar compilation). All the surveys 
cited below provide best-fit luminosity function (LF) parameters, which are then used to integrate the LF down to the 
same relative limiting luminosity, $L_{\rm min}/L_\star=0.01$. Most of these LFs have faint-end slopes $>-1.7$, which makes 
the corresponding volume emissivities rather insensitive to the value of the adopted limiting luminosity.
\citet{Schulze09} 
combined the Sloan Digital Sky Survey (SDSS) and the Hamburg/ESO survey results into a single $z=0$ AGN LF covering 4 orders 
of magnitude in luminosity. In the redshift range $0.68<z<3.0$, the $g$-band LF of \citet{Palanque-Delabrouille13} combines 
SDSS-III and Multiple Mirror Telescope quasar data with the 2SLAQ sample of \citet{Croom09}. The $1<z<4$ AGN LF by \citet{Bongiorno07} 
again merges SDSS data at the bright end with a faint AGN sample from the VIMOS-VLT Deep Survey. 
The high-redshift quasar LF in the Cosmic Evolution Survey (COSMOS) in the bins $3.1<z<3.5$ and $3.5<z<5$ has been investigated by 
\citet{Masters12}, who find a decrease in the space density of faint quasars by roughly a factor of four from redshift 3 to 4. 
A significantly higher number of faint AGNs at $z\sim 4$ is found by \citet{Glikman11} in the NOAO Deep Wide-Field Survey 
and the Deep Lens Survey, and by  \citet{Giallongo15} at $z=4-6$ in the CANDELS GOODS-South field. 
A novel detection criterion is adopted in \citet{Giallongo15}, whereby high-redshift galaxies are first selected in the NIR $H$ band 
using photometric redshifts, and become AGN candidates if detected in X-rays by {\it Chandra}. AGN candidates are found to have X-ray luminosities and 
rest-frame UV/X-ray luminosity ratios that are typical of Seyfert-like and brighter active nuclei. 
If correct, these claims suggest that AGNs may be a more significant contributor to the ionizing background radiation than previously estimated.  

We have converted the integrated optical emissivity inferred from these studies, $\epsilon_{\lambda}$ (in units of $\lumdens$),
into a 1 ryd emissivity, $\epsilon_{912}$, 
using a power-law spectral energy distribution, $\epsilon_{912}=\epsilon_{\lambda} (\lambda/912)^{-\alpha_{\rm uv}}\,\bar f_{\rm esc}$, 
with $\alpha_{\rm uv}=0.61$ following \citet{Lusso15}. We assume an escape fraction of hydrogen-ionizing radiation 
$\bar f_{\rm esc} = 1$. To assess whether a faint AGN population 
can dominate the cosmic reionization process under reasonable physical assumptions, we adopt in the following 
an AGN comoving emissivity of the form 
\begin{equation}
\log \epsilon_{912}(z)=25.15e^{-0.0026z}-1.5e^{-1.3z}, 
\label{eqemiss}
\end{equation}
for $z<z_{\rm QSO}$, and zero otherwise.
Despite the significant scatter in the data points, this function fits reasonably well the $z=0$, $z<2.5$, and $4<z<5$ emissivities 
from \citet{Schulze09}, \citet{Bongiorno07}, and \citet{Giallongo15}, respectively. Note that this emissivity does not drop 
at high redshift like, e.g., the LyC emissivity of luminous quasars inferred by \citet{Hopkins07} (see Fig. \ref{fig1}). 
It is also higher compared to previous estimates at low redshift, a fact that could contribute to solve the ``photon underproduction
crisis" of \citet{Kollmeier14} (see also \citealt{Khaire15}). 

\section{Reionization History}

Reionization is achieved when ionizing sources have radiated at least one LyC photon per atom, and the rate of LyC photon production is sufficient 
to balance radiative recombinations. Specifically, the time-dependent ionization state of the IGM can be modeled semi-analytically by integrating the 
``reionization equations" \citep{Madau99,Shapiro87} 
\begin{align}
{dQ_\nHII\over dt} & ={\dot n_{\rm ion,H}\over \langle n_\nH \rangle}-{Q_\nHII\over t_{\rm rec,H}}\\
{dQ_\nHeIII\over dt} & ={\dot n_{\rm ion,He}\over \langle n_\nHe\rangle}-{Q_\nHeIII\over t_{\rm rec,He}}
\end{align}
for the volume fractions $Q$ of ionized hydrogen and doubly-ionized helium. Here, the angle brackets denote a volume average, gas densities are expressed in comoving units, 
$t_{\rm rec}$ is a characteristic recombination timescale, and $\dot n_{\rm ion}=\int d\nu (\epsilon_\nu/h\nu)$ is the 
injection rate density of ionizing radiation, i.e. photons between 1 and 4 ryd in the case of \HI\ ($\dot n_{\rm ion,H}$) and above 4 ryd for \HeII\ ($\dot n_{\rm ion,He}$). 
We do not explicitly follow the transition from neutral to singly-ionized helium, as this occurs nearly simultaneously to and cannot be readily decoupled from 
the reionization of hydrogen. 

The ODEs above assume that the mean free path of UV radiation is always much smaller than the horizon and that the absorption of photons 
above 4 ryd is dominated by \HeII. Because only a negligible amount of recombinations occurs in mostly neutral gas,
these equations do not explicitly account for the presence of optically thick absorbers that reduce the mean free path of LyC radiation and 
may further delay reionization \citep{Bolton09}. They allow, mathematically, for values of $Q$ that are $>1$, which is physically impossible.        

\begin{figure*}[thb]
\centering
\includegraphics*[width=0.49\textwidth]{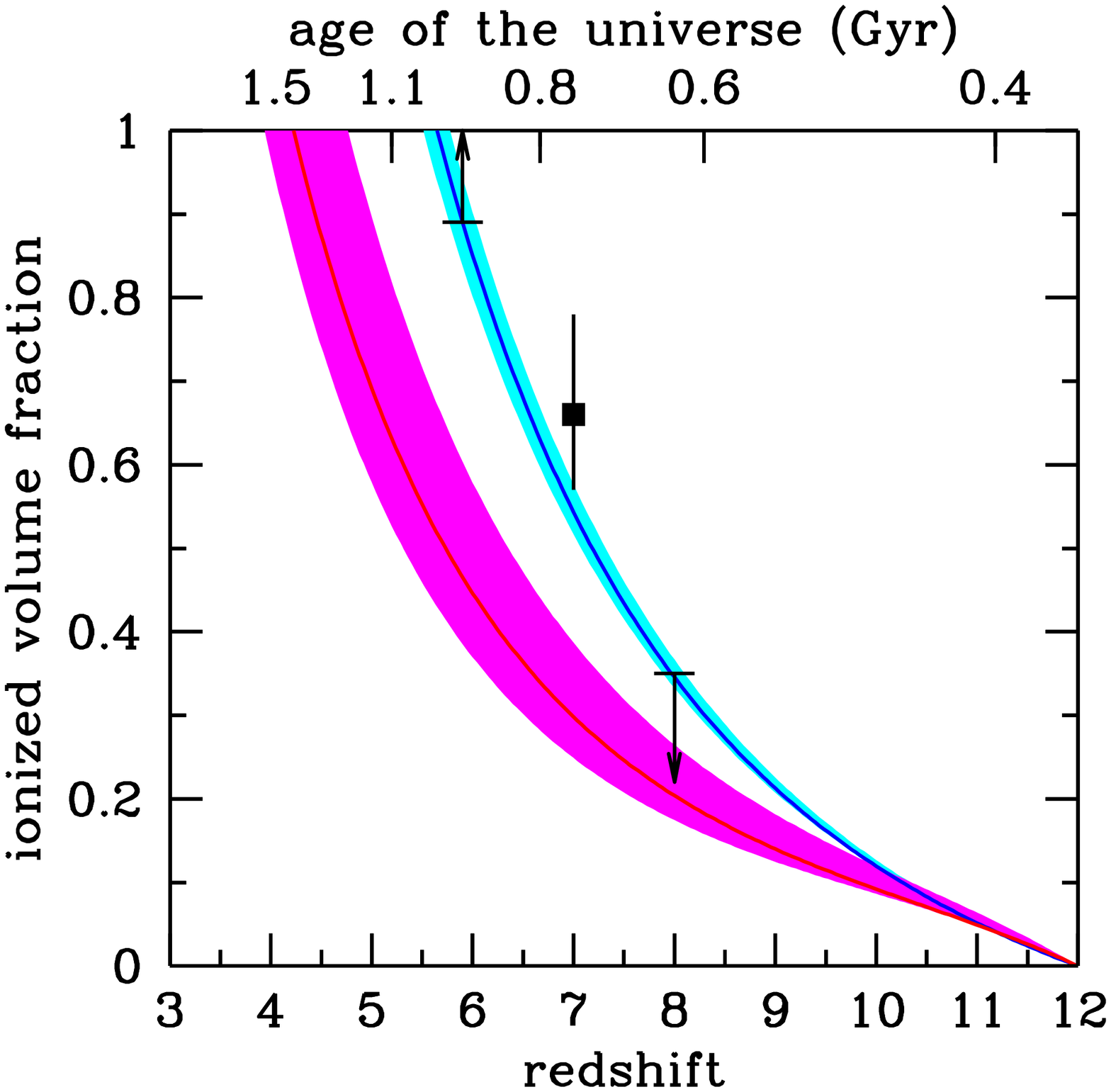}
\includegraphics*[width=0.49\textwidth]{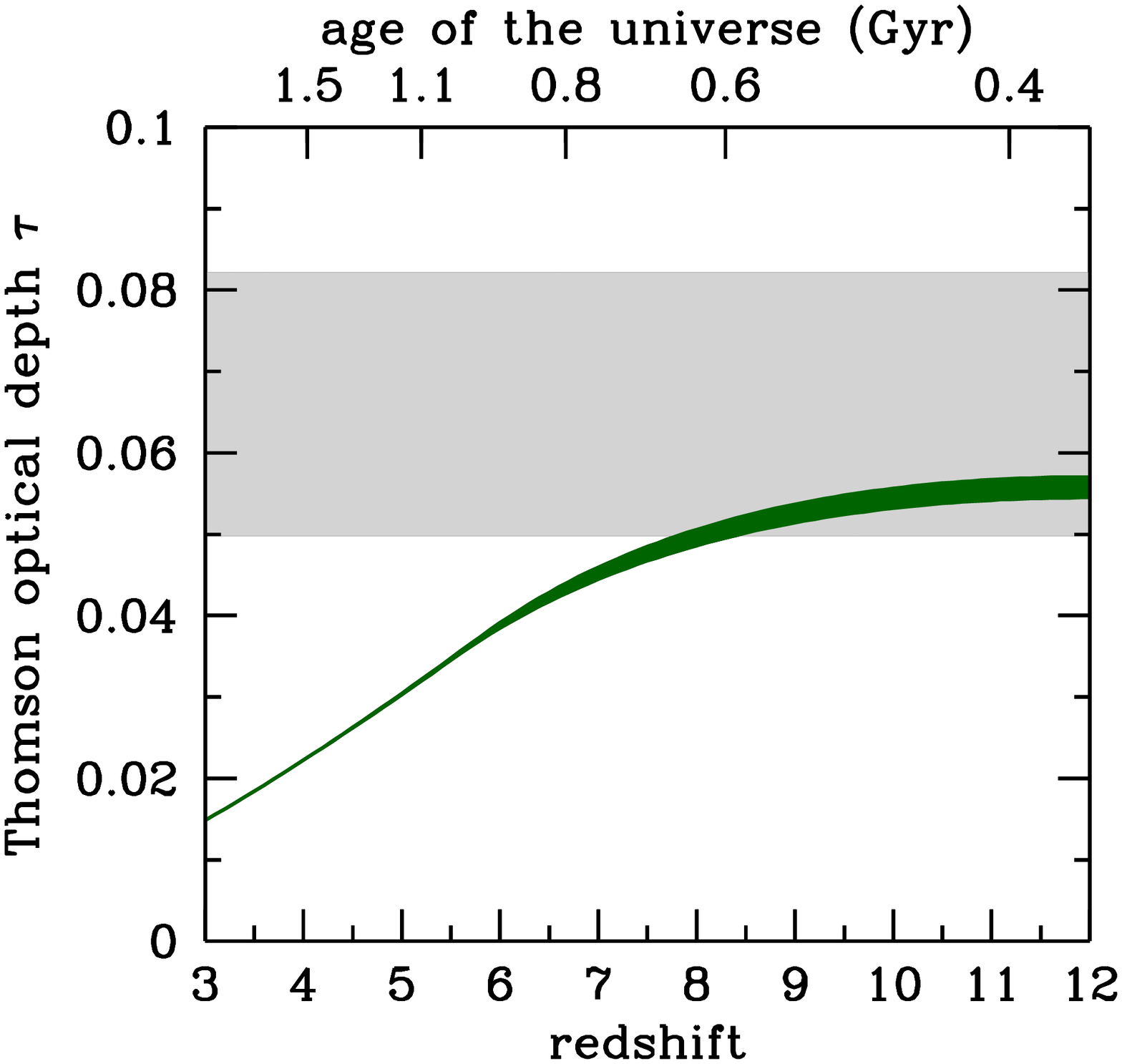}
\caption{\footnotesize Reionization history for our AGN-dominated scenario.  Left panel: evolving \HII\ (blue) and \HeIII\ (magenta) 
ionized volume fractions for our AGN-dominated scenario. Hydrogen in the IGM is fully reionized when $Q_\nHII=1$, while helium is doubly reionized
when $Q_\nHeIII=1$. The solid lines correspond to our default model parameters, while the shading shows the effects of changing clumping factor, 
IGM temperature, and EUV spectral slope (see text for details). Note the ``non-standard" early reionization of helium, $Q_\nHeIII=1$ at $z\gta 4$. 
The data point at $z=7$ and the upper limit at $z=8$ show the constraint on the neutral hydrogen fraction of the IGM inferred from the 
redshift-dependent prevalence of \Lya\ emission in the UV spectra of $z=6-8$ galaxies \citep{Schenker14}.
The 1$\sigma$ lower limit at $z=5.9$ shows the bound on the neutral hydrogen fraction of the IGM inferred from the dark pixel statistics
\citep{McGreer15}.
Right panel: Thomson optical depth to electron scattering, $\tau$, integrated over redshift from the present day (green curve). The Planck constraint 
$\tau=0.066\pm 0.016$ is shown as the gray area.
}
\label{fig2}
\vspace{+0.cm}
\end{figure*}
Following the results of cosmological hydrodynamical simulations by \citet{Shull12} (see also \citealt{Finlator12}), we define the characteristic 
hydrogen recombination timescale as 
\begin{equation}
t_{\rm rec,H}=[(1+\chi) \langle n_\nH \rangle (1+z)^3 \langle \alpha_B(T)\rangle \,C_{\rm RR}]^{-1}, 
\end{equation}
where $\alpha_B$ is the case-B radiative recombination rate coefficient, $\chi\equiv Y/[4(1-Y)]=0.083$ allows
for the presence of photoelectrons from \HeII\ (here $Y$ is the primordial helium mass fraction), and $C_{\rm RR}=2.9[(1+z)/6]^{-1.1}$ is 
the clumping factor of ionized hydrogen that accounts for both density and temperature effects on the average recombination rate. 
Similarly, the recombination timescale of doubly ionized helium is
\begin{equation}
t_{\rm rec,He}=[(1+2\chi) \langle n_\nH\rangle (1+z)^3 Z \langle \alpha_B(T/Z^2)\rangle \,C_{\rm RR}]^{-1}, 
\end{equation}
where $Z=2$ is the ionic charge and we have assumed that \HII\ and \HeIII\ have similar clumping factors. We have numerically integrated the reionization 
equations from $z_{\rm QSO}=12$ onwards, extrapolating the AGN emissivity in Equation (\ref{eqemiss}) to $z_{\rm QSO}$, and 
assuming an EUV power-law spectrum $\propto \nu^{-\alpha_{\rm euv}}$ with $\alpha_{\rm euv}=1.7$ \citep{Lusso15} and a gas temperature of $T=20,000\,$K. 
The integrated electron scattering optical depth can be calculated as 
\begin{equation}
\tau(z)=c\sigma_T \langle n_{\rm H} \rangle \int_0^{z}{(1+z')^2dz'\over H(z')}[Q_\nHII (1+\chi)+\chi Q_\nHeIII],
\label{eq:taues}
\end{equation}
where $c$ is the speed of light, $\sigma_T$ the Thomson cross section, $H(z)$ is the Hubble parameter, and we assumed $Q_\nHeII=Q_\nHII-Q_\nHeIII$. 

Figure \ref{fig2} shows the resulting ionization history, quantified by $Q_\nHII$, $Q_\nHeIII$, and $\tau(z)$. The shading shows the 
effects of changing clumping factor [$C_{\rm RR}=9.25-7.21\log(1+z)$, \citealt{Finlator12}], IGM temperature ($T=15,000\,$K), and 
EUV spectral slope ($\alpha_{\rm euv}=1.57$, \citealt{Telfer02,Stevans14}). With our default parameters, {\it hydrogen reionization is completed by 
$z\simeq 5.7$, helium is doubly ionized by $z\simeq 4.2$, and the Thomson scattering optical depth is $\tau\simeq 0.056$}. 
The last is consistent with the value reported by the Planck collaboration at the $1\sigma$ level.  
The first agrees with an updated analysis by \citet{Becker15} of the line-of-sight variance in the intergalactic \Lya\ opacity at $4<z<6$,
showing that the data near $z=5.6-5.8$ require fluctuations in the volume-weighted hydrogen neutral fraction that are higher than 
expected from density variations alone. These fluctuations are most likely driven by large-scale variations in the mean free path, 
a signature of the trailing edge of the cosmic reionization epoch. As shown in Figure \ref{fig2}, our reionization history is also consistent with 
the dark pixel fraction 
observed in the \Lya\ and \Lyb\ forest of $z>6$ quasars, which provides a model-independent upper limit of $1-Q_\nHII <0.11$ at $z = 5.9$ \citep{McGreer15}.

The redshift-dependent fraction of color-selected galaxies revealing \Lya\ emission has become a valuable constraint on the evolving neutrality 
of the early IGM. We notice here that, in our late reionization model, the hydrogen neutral fraction evolves quite rapidly from redshift 6 
($1-Q_\nHII=0.15$) to redshift 8 ($1-Q_\nHII=0.65$). This will cause a fast decline of the mean \Lya\ line transmissivity of the 
IGM and may explain the swift drop observed in the space density of \Lya\ emitting galaxies at $z>6$ \citep[e.g.,][]{Choudhury15,Schenker14}.   

While a late hydrogen reionization epoch is not a unique feature of an AGN-dominated scenario, the early reionization of \HeII\ is. Traditionally, 
the peak in luminous quasar activity observed at $z\approx 2.5$ has been expected to coincide with the end of helium reionization. 
And while observations of patchy absorption in the \HeII\ \Lya\ forest at these epochs have been interpreted as signaling the tail end of 
\HeII\ reionization at $z\le 2.7$ \citep[e.g.,][and references therein]{Shull10}, the large fluctuations in the background radiation field above 4 ryd 
predicted from the rarity of bright quasars and the relatively short attenuation lengths of EUV photons make the interpretation of the 
\HeII\ data still controversial. Indeed, rather than complete Gunn-Peterson absorption at higher redshifts, recent observations have revelead high-transmission 
regions out to $z=3.5$ \citep{Worseck15}, in conflict with numerical models of \HeII\ reionization driven by luminous quasars 
\citep{McQuinn09,Compostella13}. According to \citet{Worseck15}, the observed mild evolution with redshift of \HeII\ absorption demands that the bulk of 
intergalactic helium was already doubly ionized at $z>3.5$ by a population of early EUV sources. The high AGN comoving emissivity present at $z>4$ 
in our model may accomplish just that. We find that singly- and doubly-ionized helium contribute about 13\% to $\tau$, and the \HeIII\ fraction is 
already 50\% when hydrogen becomes fully reionized at redshift 5.7.   

\begin{figure}[thb]
\centering
\includegraphics*[width=0.49\textwidth]{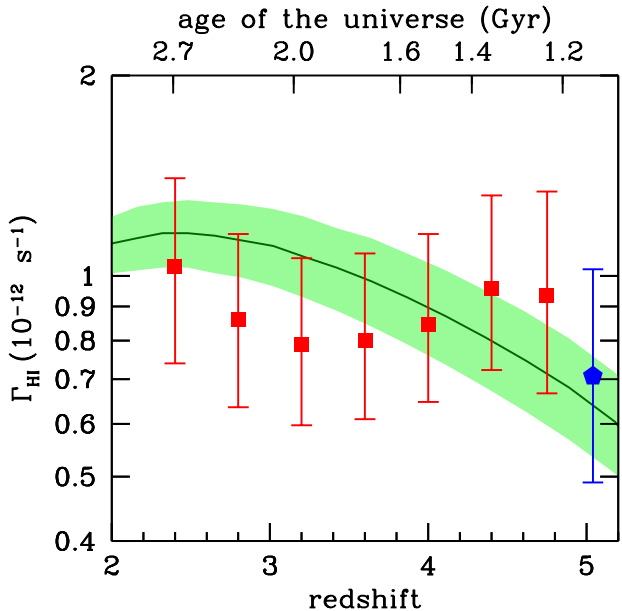}
\caption{\footnotesize The hydrogen photoionization rate, $\Gamma_\nHI$, from $z=2$ to $z=5$. The green solid line corresponds to our default AGN-dominated model, 
while the shading shows the effects of changing the IGM mean free path to ionizing radiation by $\pm 25$ percent.
{Red squares:} empirical measurements from the \Lya\ forest effective opacity by \citet{Becker13}. {Blue pentagon:} same 
using the quasar proximity effect \citep{Calverley11}.
}
\label{fig3}
\vspace{+0.0cm}
\end{figure}

\section{Discussion}

Only the most luminous distant quasars can be detected in surveys such as the SDSS, leaving the contribution of faint AGNs to the early ionizing 
background highly uncertain. Recent multiwavelength deep surveys have suggested the presence of a hitherto unknown population of faint AGNs 
\citep{Fiore12,Giallongo15} and have prompted a reexamination of the role played by AGNs in the reionization of the IGM.
We have expanded on previous studies and assessed a model in which the UV radiation responsible for the reionization of hydrogen and helium 
in the universe arises entirely from quasars and active galaxies. We have assumed here that normal star-forming galaxies will make only a negligible 
contribution to the AGN LyC emissivity given in Equation (\ref{eqemiss}), which is true provided their globally-averaged escape fraction does 
not exceed a few percent or so \citep[e.g.,][]{Madau14}. 

Compared to the standard picture widely discussed in the literature, the AGN-dominated scenario examined in this work completes 
hydrogen reionization late ($z\lta 6$), double helium reionization early ($z\gta 4$), and produces a low electron scattering optical depth 
that is consistent with the Planck value at the $1\sigma$ level. It may provide an explanation to some 
otherwise puzzling recent findings, from the rapid decline of the space density of \Lya\ emitting galaxies observed at $z>6$, to an IGM whose 
temperature is found to increase from redshift 5 to 2 and where \HeII\ appears to be predominantly ionized at $z\simeq 3.5$. It is, of course, a 
model that is also plagued by a number of uncertanties regarding the properties of faint AGNs (their space densities and EUV spectra at high-$z$), 
and of the early IGM (its temperature and clumpiness).  As already pointed out by \citet{Giallongo15}, a critical assumption 
is the high escape fraction of ionizing radiation needed for the global AGN population to dominate reionization.     
While no discernible continuum edge at 912 \AA\ is seen in a composite FUV spectrum of 159 $0<z<1.5$ quasars and active galaxies obtained with the 
Cosmic Origins Spectrograph (COS) \citep{Stevans14}, it is uncler whether escape fractions of order unity are also typical of fainter, higher redshift AGNs.
The model is also somewhat sensitive to $z_{\rm QSO}$, the ``formation epoch'' of the earliest AGNs.
Here, we have assumed $z_{\rm QSO}=12$ and a comoving AGN emissivity that is only a factor of 3 smaller at $z_{\rm QSO}$ than at redshift 4.      
In a model with $z_{\rm QSO}=9$, for example, hydrogen reionization would only be completed by $z\simeq 5.4$.

As a corroboration of the slowly evolving LyC emissivity adopted in this work, we have run a modified version of our radiative transfer
code CUBA \citep{Haardt12} to estimate the intensity and spectrum of the filtered ionizing background -- specifically the hydrogen photoionization 
rate $\Gamma_\nHI$ -- predicted by our AGN-dominated scenario,          
\begin{equation}
\Gamma_\nHI\equiv \int_{\nu_\nHI}^\infty {d\nu\,\frac{4\pi J_\nu}{h\nu}\sigma_\nHI(\nu)}. 
\end{equation}
Here, $J_\nu$ is the angle-averaged monochromatic intensity, $h$ is the Planck constant, and $\sigma_\nHI$ and $\nu_\nHI$ are the hydrogen photoionization 
cross section and ionizing threshold frequency. We have computed the 
background ionizing intensity $J$ using the emissivity in Equation (\ref{eqemiss}), a EUV spectrum with $\alpha_{\rm euv}=1.7$, and a new a piecewise parameterization of the
distribution in redshift and column density of intergalactic absorbers that fits the measurements of the mean free path of 1 ryd photons by \citet{Worseck14}.
The results, shown in Figure \ref{fig3}, are in formal agreement with empirical determinations of $\Gamma_\nHI$ in the interval $2<z<5$ based on the \Lya\ opacity of the 
IGM \citep{Becker13}. Note that, in the traditional view, the slowly evolving emissivity needed to reproduce the \Lya\ opacity data can only be achieved 
by carefully tuning the escape fraction from star-forming galaxies to increase rapidly with lookback time, so as to compensate for the decline in the 
star formation activity towards early epochs \citep[e.g.,][]{Haardt12,Kuhlen12}.

\begin{figure}[thb]
\centering
\includegraphics*[width=0.49\textwidth]{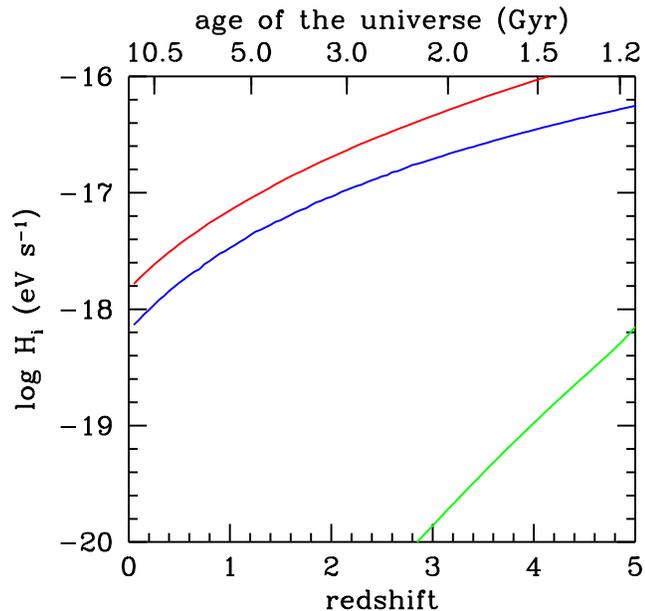}
\caption{\footnotesize Photoheating rates for \HI\ (blue curve), \HeI\ (green curve), and \HeII\ (red curve) in our AGN-dominated model, from the present epoch 
to $z=5$, i.e. after the reionization of hydrogen. All photoheating rates are expressed per hydrogen atom in units of eV s$^{-1}$, and are computed for gas 
at the mean cosmic density.
}
\label{fig4}
\vspace{+0.0cm}
\end{figure}

In our AGN-dominated scenario, the heat input from helium reionization will start dominating the thermal balance of the IGM earlier than in the standard picture. 
In Figure \ref{fig4} we plot the photoheating rates after the reionization of hydrogen for gas at the mean density, 
\begin{equation}
H_i= \langle n_i/n_\nH \rangle \int_{\nu_i}^\infty J_\nu \sigma_i (h\nu-h\nu_i)d\nu/(h\nu), 
\end{equation}
for species $i=$ \HI, \HeI, or \HeII. Here, $\sigma_i$ and $\nu_i$ are the photoionization cross section and threshold frequency for the respective species, and $n_i$ 
is their abundance (computed integrating the non-equilibrium rate equations for gas at the mean density). The formula above should provide the correct mean heating
rate of intergalactic gas once the background intensity $J_\nu$ is properly reprocessed while propagating through the IGM \citep{Puchwein15}. 
\HeII\ photoheating exceeds the hydrogen term by nearly a factor of three. Measurements of the IGM temperature evolution from redshift 5 to 2 derived from the \Lya\ forest 
have been known to be inconsistent with the mononotic decrease with redshift expected after the completion of hydrogen reionization \citep{Becker11}. 
This discrepancy may be solved by the additional heating provided by an earlier and more extended period of \HeII\ reionization.

The population of faint, high redshift AGNs invoked here should leave an imprint on the cosmic X-ray background (XRB). 
\citet{Moretti12} derived a value of $J_{\rm 2\,keV}\simeq 2.9^{+1.6}_{-1.3}\times 10^{-27}\,\uvunits$ to the $2\,$keV unresolved XRB. The expected contribution 
at 2 keV from AGNs above redshift $z_x$ can be estimated as \citep{Haardt15} 
\begin{equation}
J_{\rm 2\,keV}={c\over 4\pi} \int_{z_x}^\infty {dz\over (1+z)H(z)}\epsilon_{\rm 2\,keV}(z) (1+z)^{-\alpha_x},
\end{equation}
where the specific comoving emissivity at 2 keV is related to the LyC emissivity by  
\begin{equation}
\epsilon_{\rm 2\,keV}(z)=\epsilon_{912}(z) \times R_{\rm II}\left({2500\over 912}\right)^{\alpha_{\rm uv}}\left({2\,{\rm keV}\over 2500\,{\rm \AA}}\right)^{-\alpha_{\rm ox}}. 
\end{equation}
Here, $\alpha_{\rm ox}$  is the optical-to-X-ray spectral index needed to join the emissivity at 2500\,\AA\ with that at 2 keV, and $R_{\rm II}$ is a correction factor that 
accounts for the possible contribution of UV obscured (``Type II") AGNs at $z\ge z_x$ to the observed XRB. Using $\alpha_{\rm uv}=0.61$ as before, $\alpha_{\rm ox}=1.37$ \citep{Lusso10}, 
$\alpha_x=0.9$, $z_x=5$, and $R_{\rm II}=2$ \citep{Merloni14} we obtain $J_{\rm 2\,keV}=1.64\times 10^{-27}\,\uvunits$, i.e. a contribution of nearly 60\% to the unresolved XRB. 
We find that $z>5$ active galaxies can reionize the universe without overproducing the unresolved XRB provided their properties (i.e. fraction of obscured objects, 
optical-to-X-ray spectral indices) are similar to those of their lower redshift counterparts \citep[cf.][]{Haardt15,Dijkstra04}. 

Finally, we conclude by pointing out that, in order to promote further testing of this model against new and old observations, we will make the results of our radiative transfer 
calculations of an AGN-dominated UV background freely available for public use at \url{http://www.ucolick.org/~pmadau/CUBA}.

\acknowledgments
\ni We thank E. Giallongo. Z. Haiman, E. Lusso, and A. Meiksin for helpful discussions on various aspects of this paper.
P.M. acknowledges support by the NSF through grant AST-1229745 and NASA through grant NNX12AF87G.

{}

\end{document}